\documentclass{article}

\PassOptionsToPackage{numbers, compress}{natbib}

\usepackage{float} 

\usepackage[final]{neurips_2020}

\usepackage[utf8]{inputenc} 
\usepackage[T1]{fontenc}    
\usepackage{hyperref}       
\usepackage{url}            
\usepackage{booktabs}       
\usepackage{amsfonts}       
\usepackage{nicefrac}       
\usepackage{microtype}      
\usepackage{amsmath}
\usepackage[numbers]{natbib}
\usepackage{mathtools, nccmath}

\usepackage{graphicx, color}

\usepackage[ruled,vlined]{algorithm2e}
\usepackage[compatible]{algpseudocode}
\renewcommand{\algorithmiccomment}[1]{\bgroup\hfill\tiny//~#1\egroup}
\title{A Scalable Inference Method For Large Dynamic Economic Systems}

\author{%
  Pratha Khandelwal\thanks{pratha.khandelwal19@imperial.ac.uk} \\
  Department of Computing\\
  Imperial College London\\
   \And
   Philip Nadler \\
   Data Science Institute \\
   Imperial College London \\
   \And
   Rossella Arcucci\thanks{r.arcucci@imperial.ac.uk} \\
   Data Science Institute \\
   Imperial College London \\
    \And
  William Knottenbelt \\
  Department of Computing \\
  Imperial College London \\
  \And
  Yi-Ke Guo \\
  Data Science Institute \\
  Imperial College London \\
}

\begin{document}

\maketitle

\begin{abstract}
The nature of available economic data has changed fundamentally in the last decade due to the economy's digitisation. With the prevalence of often black box data-driven machine learning methods, there is a necessity to develop interpretable machine learning methods that can conduct econometric inference, helping policymakers leverage the new nature of economic data. 
We therefore present a novel Variational Bayesian Inference approach to incorporate a time-varying parameter auto-regressive model which is scalable for big data. 
Our model is applied to a large blockchain dataset containing prices, transactions of individual actors, analyzing transactional flows and price movements on a very granular level. The model is extendable to any dataset which can be modelled as a dynamical system. We further improve the simple state-space modelling by introducing non-linearities in the forward model with the help of machine learning architectures.
\end{abstract}

\vspace{-0.7em}
\section{Introduction}

\vspace{-0.5em}
The digitisation of the economy and society had a profound impact on how researchers study economic interactions. The availability of enormous granular datasets has led to innovations in data-driven machine learning approaches, excelling in forecasting. However, the opaqueness of these approaches has been proven to limit the usefulness to policymakers. We thus develop a scalable inference model which combines innovations of popular machine learning methods such as variational approaches and non-linear forecasting methods with interpretable parametric models to study economic interactions at scale. We apply the model to study an emerging part of this new digital economy. \\
A prime example of this digitisation is the introduction of decentralised digital currencies which had a big impact on the global financial system. Blockchain, by using decentralised architecture, has been a disruptive innovation in terms of global transactions. With increasing interest in cryptocurrency trading, especially in ERC20 tokens, and increase in the number of exchanges providing a  platform for the users to trade with ease, cryptocurrency trading has now become a complex economic system with complicated dynamics between price, trading volume and tokenflow.\\
Due to a large number of transactions and with every transaction being accessible online and traceable to some extent \cite{10.1257/jep.29.2.213}, we are left with granular data which can be used for extracting valuable new information about this ecosystem, like, interactions between on-chain (transaction amount and token inflow) and off-chain (trade data like closing price) ecosystems. Both these features being complementary, create a complex dynamical system similar to what we see in traditional financial markets. This granular dataset, in combination with our model, provides a data-driven approach to inform policymakers and can provide valuable information when creating new regulations to protect investors in this yet largely unregulated market.  \\
Previous works have successfully modelled cryptocurrencies as a dynamical system using Time-Varying Parameter - Vector Autoregression (TVP-VAR) \cite{hotz2018predicting,nadler2019data}. They note that a time-varying regression coefficient could ideally model this ecosystem. As noted by \citet{nadler2019data} and \citet{haldane2018will}, the economic model needs to scale to deal with big data appropriately. We build on the previous research and improve the Bayesian inference in a way that is scalable for big data. We introduce a novel algorithm, Time-Varying Parameter - Vector Autoregression - Variational Inference (TVP-VAR-VI) to tackle this problem. We test the proposed approach on synthetic data to validate and perform benchmarking. We then extended it to real-world blockchain data. We use this inference technique to interpret an otherwise unobservable parameter representing the dynamics of the blockchain ecosystem. The resulting parameter vector in latent space is used to visualise and interpret the influence of on-chain transactions on trading data, pricing and market dynamics. We use Bayesian inference as opposed to a neural network, as a simplistic NN forecasting architecture would yield a uninterpretable weight matrix in latent space. The parameter estimated using our technique can be interpreted visually, which has been further discussed in Section \ref{exp}.\\
We also challenge the simplistic state-space model for the evolution of the latent parameter. We propose a new technique to improve the state-space modelling of the ecosystem, by incorporating non-linearity in our modelling by using data-driven approaches and develop a neural network architecture called TVP-VARNet.\\
In short, the Bayesian inference model helps us in obtaining the latent dynamic parameter which describes the interrelationship of economic variables dynamically over time whose structure economists can exploit for policy analysis. Our methodology is especially relevant for very large and granular datasets which would be computationally prohibitive for many established econometric models, since our model is a scalable algorithm for big data. Extending TVP-VAR-VI, we develop TVP-VARNet model which helps in overcoming the shortcoming of the simplistic state-space to forecast this latent parameter in case of out-of-sample n-step predictions.


\section{Background} 
\vspace{-0.5em}
In the following, we will denote with $x_t \in \mathbb{R}{^n}$, the time evolving latent variables and ${y_t} \in \mathbb{R}{^m}$ as observations.  $\mathcal{M}$ denotes the operator used to evolve the latent variables in time
        \begin{equation}\label{stateEq}
            {x_{t}} = \mathcal{M}({x_{t-1}}) + \epsilon_t
        \end{equation}
        where $\epsilon_{t}$  is Gaussian background error with covariance matrix ${Q}$. Finally, $\mathcal{H}$: $\mathbb{R}{^n}\to\mathbb{R}{^m}$ will denote the operator which maps the state variables to the observations.
        \begin{equation} \label{stateEq2}  {y_{t}} = \mathcal{H}( {x_{t}}) + \epsilon_o(t) \end{equation}
    where $\epsilon_o(t)$ is Gaussian model error with with covariance matrix $ {R}$.


\subsection{Bayesian Inference}
\vspace{-0.5em}
The objective of Bayesian inference is to update the belief of latent space (analysis, $ {x^a_t}$) by ingesting the observation ($ {y_t}$) and prior/model information ($ {x^b_t}$)  at each time step, t. 
The prior distribution here is the PDF of the latent state when no observations are available, $ {x_t^b}$ (i.e. output of the model \eqref{stateEq}). The likelihood is the observations we get using the latent space in \eqref{stateEq2}. Whereas, the posterior represents the final updated latent state, after considering observations \cite{bannister}. Modelling these PDFs as gaussian process, with prior mean as $ {x_t^b}$, covariance Q and likelihood mean as $\mathcal{H}( {x_{t}})$, covariance R, we get 
\begin{equation}
\label{eq:cost}
P( {x_t}| {y_t}, {x_t^b}) \propto \exp -\frac{1}{2}
\big[( {x_t}- {x^b_t})^{\mathrm{T}} {Q_t}^{-1}( {x_t}- {x^b_t}) + ({y_t}-\mathcal{H}( {x_t}))^{\mathrm{T}}\ {R_t}^{-1}( {y_t}-\mathcal{H}( {x_t})) \big]
\end{equation}
We seek a solution maximising the posterior probability, $P( {x_t}| {y_t}, {x_t^b})$ which means minimising the negative log likelihood, $- \ln P( {x}| {y}, {x_b})$ or $J( {x^a_t})$
\begin{equation} \label{eq1}
\begin{split}
J( {x^a_t}) &= ( {x_t}- {x^b_t})^{\mathrm{T}} {Q_t}^{-1}( {x_t}- {x^b_t}) + ( {y_t}-\mathcal{H}( {x_t}))^{\mathrm{T}} {R_t}^{-1}( {y_t}-\mathcal{H}( {x_t})) \\
&=  {J_b(x_t) + J_o (x_t)}
\end{split}
\end{equation}
Proven techniques for such inference schemes are the Kalman filter and Variational inference.
The Kalman filter \cite{kalman1960general} approaches the problem statement sequentially by analytically solving the cost function and calculating a weighted average, known as Kalman Gain. In contrast, the Variational inference method minimises the same cost function that reduces the analysis gap between both time distributed observations and model solution \cite{aschDataAssimilationMethods2016}, thus working in a continuous way.

In the later sections, we will be reporting results of both Kalman and Variational inference methods, modified and implemented to suit TVP-VAR modelling. This is further extended to the blockchain data to understand the latent state dynamics.
\vspace{-0.4em}
\subsection{Ethereum and ERC20 tokens}
\vspace{-0.6em}
Ethereum is a specialised decentralised network that, along with recording transactions on blockchains, allows the creation of smart contracts. This gives an environment to create decentralised applications (DAPPs). A detailed description of Ethereum's architecture in available in \citet{wood2014ethereum}.

\textbf{ERC20 Tokens} An Ethereum Token is a digital, blockchain-based asset created on top of the Ethereum network created using a smart contract. These tokens serve as proof of ownership of an amount of gold or a house. 
\begin{figure}[h!]
    \centering
    \includegraphics[scale=0.23]{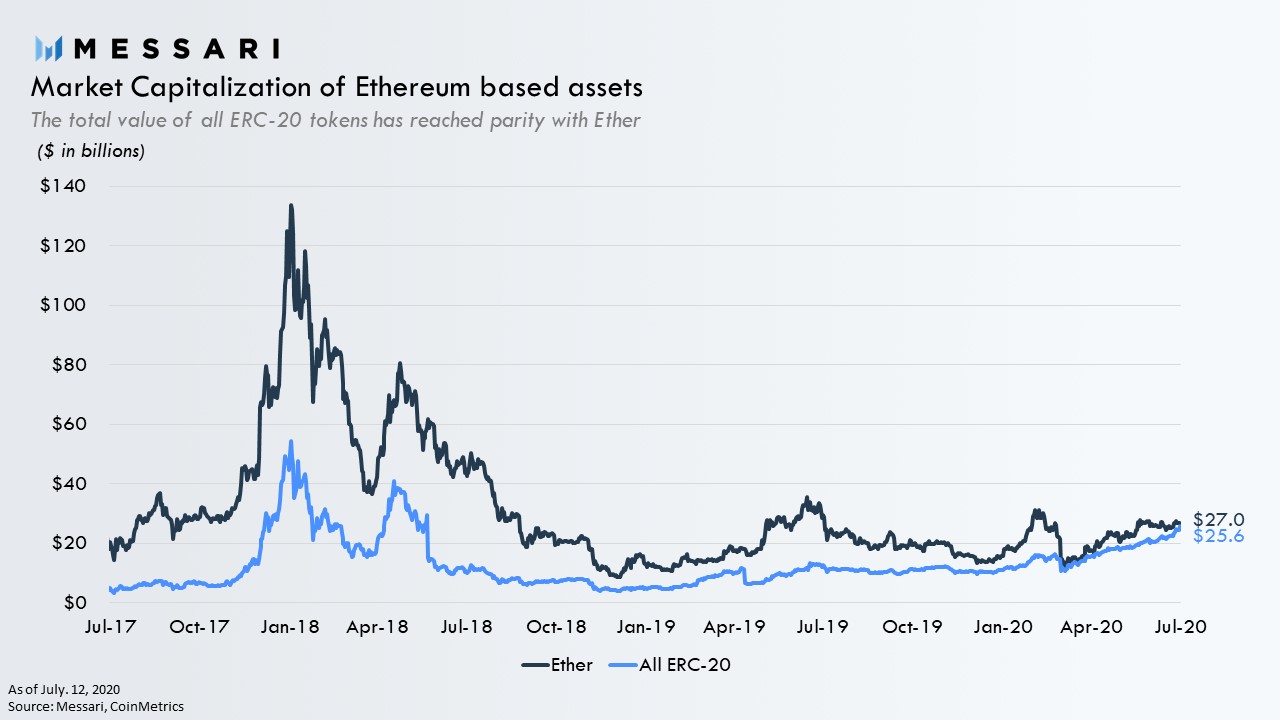}
    \caption{Market Capitalisation, Source: Messar \cite{messari_crypto_news}}
    \label{fig:messari}
\end{figure}
Figure~\ref{fig:messari} revealed that the market capital of ERC-20 tokens now represents 49\% of total assets on Ethereum. This increased share incentivizes economic activity using ERC20 tokens, providing excellent avenues for researching the trading activities and dynamics existing within this ecosystem.

\textbf{Exchanges} Cryptocurrency exchanges are online platforms where customers can trade one kind of digital asset or currency for another based on the market value of the given assets. These exchanges are intermediary between buyers and sellers of the cryptocurrency similar to the traditional stock exchange platforms. The most popular exchanges are currently Binance, GDAX, Poloniex etc \cite{frankenfield_2020}. Hence, we can see that these exchanges serve an integral part of the trading ecosystem in blockchain. These exchanges together process nearly \$10 billion \cite{blockgeeks_blockgeeks_2020}, playing a major role in driving this ecosystem.

\section{Blockchain Ecosystem: Econophysical Analysis}
\vspace{-0.5em}
\subsection{Dynamic system formulation}
We model the on-chain (transaction flows and amounts) and off-chain trading activities on blockchain exchanges as a Time Varying Parameter Vector- Autoregressive model (TVP-VAR).\\
TVP-VAR embodies a system which has a set of vector autoregressive coefficients of time-varying nature. TVP-VAR has been used to model price dynamics of various cryptocurrencies in other literature \cite{blockchainTVP} as well. The dynamic relation in the blockchain ecosystem can be time varying and thus TVP VAR helps us in accommodating this shift by varying coefficients, allowing the model parameters to vary across time. The multivariate lagged VAR is 
\begin{equation*}
    y_t = \phi_1 y_{t-1} + ... + \phi_l y_{t-l} + \mu_t + \epsilon_t \sigma_t 
\end{equation*}
where $y_t$ is $q \times 1$ vector of observations, $\mu_t$ a vector of means and $\phi_l$ is a $q \times q$ coefficient matrix and $l$ is the lag length. The model can be expressed in compact notation using $X_t= [y_{t-1}' , ..., y_{t-l}' , 1]'$ and $\phi = [\phi_1, ..., \phi_l, \mu ]'$. With $K = (ql+1)$, we get $X_t$ of dimension $K 
\times 1$ and $\phi$ as $K \times q$. We further define $x_t= I \otimes X_t $ using Kronecker product $\otimes$ and $I$ as $q \times q$ identity
matrix and define $\beta_t = vec(\phi)$, $vec()$ stacking the columns of a matrix, dim $Kq \times 1$. This allows
us to rewrite the TVP-VAR in compact notation. Further details can be found in \cite{nakajima2011time,nadlerScalableApproachEconometric2019}: 
\begin{align}
	y_t = x_t' \beta_t + \sigma \label{state_link} \\
	\beta_t = F\beta_{t-1} + v_t  \label{state2_link}
\end{align}
Where $v_t$ and $\sigma$ are zero mean error co-variances with $ v_t \sim N(0,Q_t)$ and $\sigma \sim N(0,R)$.
$\beta_t$ is the time-varying coefficient or the latent parameter that define the dynamic of the economic system. $F$ is the forward model, similar to $M$ in \eqref{stateEq} to evolve $\beta$ in time.
\vspace{-0.9em}
\subsection{Link to Bayesian inference}
\vspace{-0.6em}
A lot of times, the state representing interrelationship may evolve with time, however, this latent state can be unknown and not directly observable. These have to be inferred from the measurements/observations, which may be very sparse and noisy. Bayesian inference can help in analysing this latent space.\\
TVP-VAR with Bayesian inference is used in estimating the latent blockchain dynamics, $\beta$ in \eqref{state_link}. This method enables the use of both sparse and highly aggregated data effectively to infer dynamics to analyse the effect of trading actions on blockchain on-chain activities or how token flow effect the pricing actions on exchanges.\\
We can see that the \eqref{state_link} and \eqref{state2_link} relate to the state-space \eqref{stateEq} and \eqref{stateEq2} discussed above.
Tweaking the state-space equations can give us a TVP-VAR formulation of our econophysical system and hence modified inference approaches can help us determine the unobserved state variable, i.e. gives us insight to the interrelationship between economic variables like $x$ and $y$.

\vspace{-0.7em}
\section{Our proposed TVP-VAR-VI:  Variational Inference} \vspace{-0.7em}
Native variational data inference needs to be adapted for modelling the state-space economic time-varying model to incorporate observations and interpret the unobserved model parameter, $\beta$. We propose a new methodology to combine TVP-VAR with variational inference, formulating a novel algorithm called TVP-VAR-VI. Rewriting \eqref{eq1} compactly and adapting it to be inline with TVP-VAR equations \eqref{state_link},\eqref{state2_link} we get a cost function that reduces the analysis gap between time distributed observations and model solution.
\vspace{-0.7em}
\begin{equation}
\begin{gathered}
\label{min_var}
    J(\beta_t) = ||\beta^b_t - \beta_{t}||_{Q^{-1}_t} + \sum_{j=t}^{j=t+\Delta t} ||y_{j} - X_{j}\beta_j||_{\sigma^{-1}}, \\
    \beta^{opt}_t = \argmin_{\beta_t} J(\beta_t)
\end{gathered}
\end{equation}
We minimise the cost function in \eqref{min_var} so as to incorporate observations, $y$ and exogenous input $X$,  to infer latent parameter $\beta$. Further details can be found in previous work by \citet{nadler2019data}.\\
In \eqref{min_var}, following notations are used:
\vspace{-0.5em}
\begin{itemize}[noitemsep]
    \item  $\Delta t$ is window size, forecasts from this time window are used for optimising the initial value.
    \item 4D Variational approach is performed over a time window $(t,t+\Delta t)$, where $\beta^{opt}_t$ is the optimised initial beta value for start of the window at t. Optimising \eqref{min_var} gives us $\beta^{opt}$ every $\Delta t$ time steps.
    \item F, model forecast is identity, to evolved $\beta^{opt}$ as a random walk process. We later try to learn the F operator, i.e evolution of beta using deep learning.
    \item $\beta^b$ is the background/prior beta, updated to $\beta^{opt}$ as observations are ingested.
\end{itemize}
\vspace{-0.6em}
When $\Delta t = 0$, the cost function becomes that of 3D-Var (as time window, the 4th dimension in 4D no longer exists) and it minimises the function considering observations only that time. We propose the following algorithm combining TVP-VAR with variational approach.

\begin{algorithm}[H]
\begin{algorithmic}[1]
\STATE{Initialisation $\beta^b_0, Q_0, \sigma$}
\FOR{t = range(0, T, $\Delta t$)}\COMMENT{t takes values like 0, $\Delta t, 2\Delta t..$}
    \STATE{$\beta_t = \argmin J(\beta)$} \COMMENT {minimise \ref{min_var} using a routine}
    
    \FOR{i = 1..$\Delta t$} 
    
        \STATE{$\beta_{t+i} =F \beta_{t+i-1}$}
    
        \STATE{$Y^{hat}_{t+i} = 
        x_{t+i}\beta_{t+i}$}
    
    \ENDFOR
    
    \STATE{$\beta^b_t = \beta_{t+\Delta t}$}
\ENDFOR
\end{algorithmic}
\caption{TVP-VAR-VI}
\label{algo1}
\end{algorithm}

As we see, there is no update step similar to Kalman (calculation of Kalman gain matrix); beta is inferred using iterative minimisation of the cost.\\
It is important to note that we need to update the background beta, $\beta^b$ to optimal $\beta^{opt}$ obtained from the minimisation and use this updated $\beta^b$  as a starting point prior for the optimiser. Convergence is slower if a random starting point prior is used.
\vspace{-0.5cm}
\subsection{Implementation details}
\vspace{-0.4em}
A challenging part of implementing TVP-VAR-VI was ﬁnding a way to computationally minimise the function in \eqref{min_var}. 
We used automatic differentiation provided by TensorFlow. By calculating the function gradient from a graph, it allows us to free ourselves from deriving the gradient by hand, which is often infeasible and complex.\\
We empirically observed that Limited-memory Broyden–Fletcher–Goldfarb–Shanno (L-BFGS) \cite{nocedal_2006_numerical}  optimiser worked best for us, LBFGS converged much faster (on an average 14\% reduction in execution time) and over performed prevalent first order optimisers like Adam, SGD by reducing Mean squared error (MSE) in order of magnitude of 5.\\
L-BFGS, being one of the effective Quasi-Newton methods, gives us the ability to incorporate second order information (by estimating the Hessian) and hence could be the reason for optimal and better convergence given its ability to represent complex structures effectively.

\vspace{-0.6em}
\section{Experiments} \label{exp}
\begin{figure}[h!]
    \centering
    \includegraphics[scale=0.35,width=\linewidth]{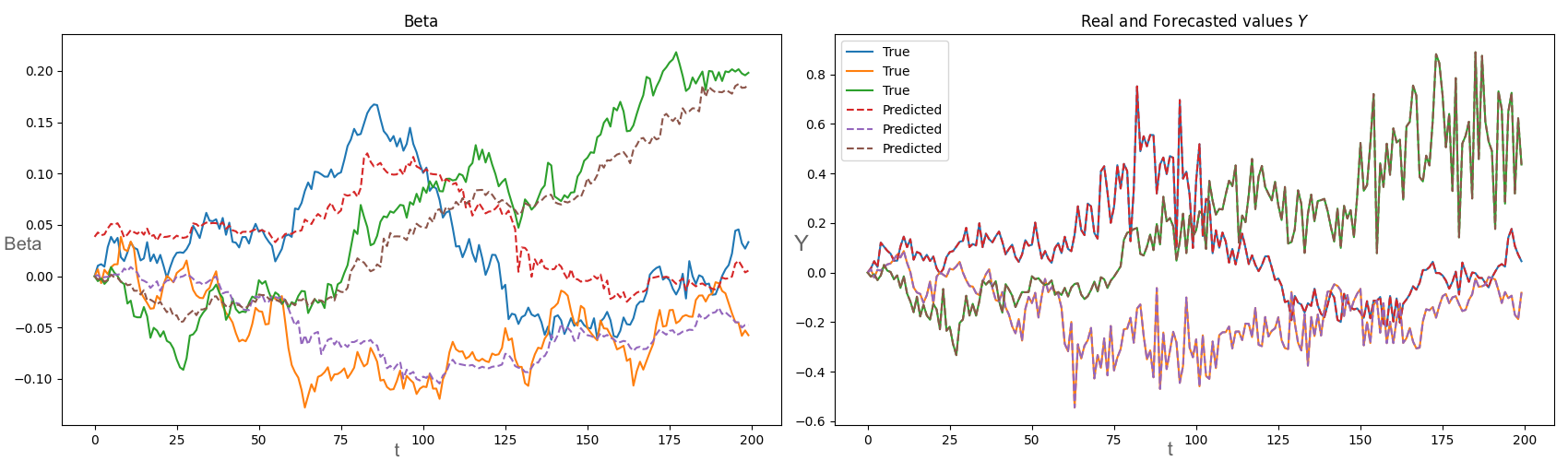}
    \caption{Window size 1, \textbf{TVP-VAR-VI}}
    \label{fig:tensorflow.png}
\end{figure}

\vspace{-0.3cm}
\paragraph{Synthetic data}
We start with a well-behaved  TVP-VAR synthetic data to evaluate the accuracy and scalability of the two approaches for big data as it gives us the ground truth for the latent parameters to validate our algorithms on. We then use the variational inference for off-chain and on-chain blockchain dataset after performing benchmarking. The data generation is similar to the one in \cite{nadlerScalableApproachEconometric2019},
\begin{equation}\label{var}
\begin{aligned}
y_t = X'_t\gamma_t + e_{1,t} \\
\gamma_t = \gamma_{t-1} + e_{2,t} + e_{3,t}
\end{aligned}
\vspace{-0.25cm}
\end{equation}
where $\gamma$ represents the time varying state variable ($\beta$), and the errors, $e_{1..3}$ are assumed to be Gaussian white noise.
Data matrix $X_t$ is generated as standardised i.i.d process. Note that the forward model operator for $\gamma$ is identity here. Figure~\ref{fig:tensorflow.png}  shows  the  sample  beta and  forecast  plots  estimated by VP-VAR-VI, i.e. our Variational approach.
We use the Mean square forecast error as our assessment metric. We implemented both the algorithms, TVP-VAR - Kalman and our novel TVP-VAR-VI using TensorFlow-gpu. This is done to ensure that the time difference noted was not due to difference in implementation details and technology used. Table \ref{benchmar} shows the results for both approaches.
\begin{table}[H]
\centering
\includegraphics[scale=0.6]{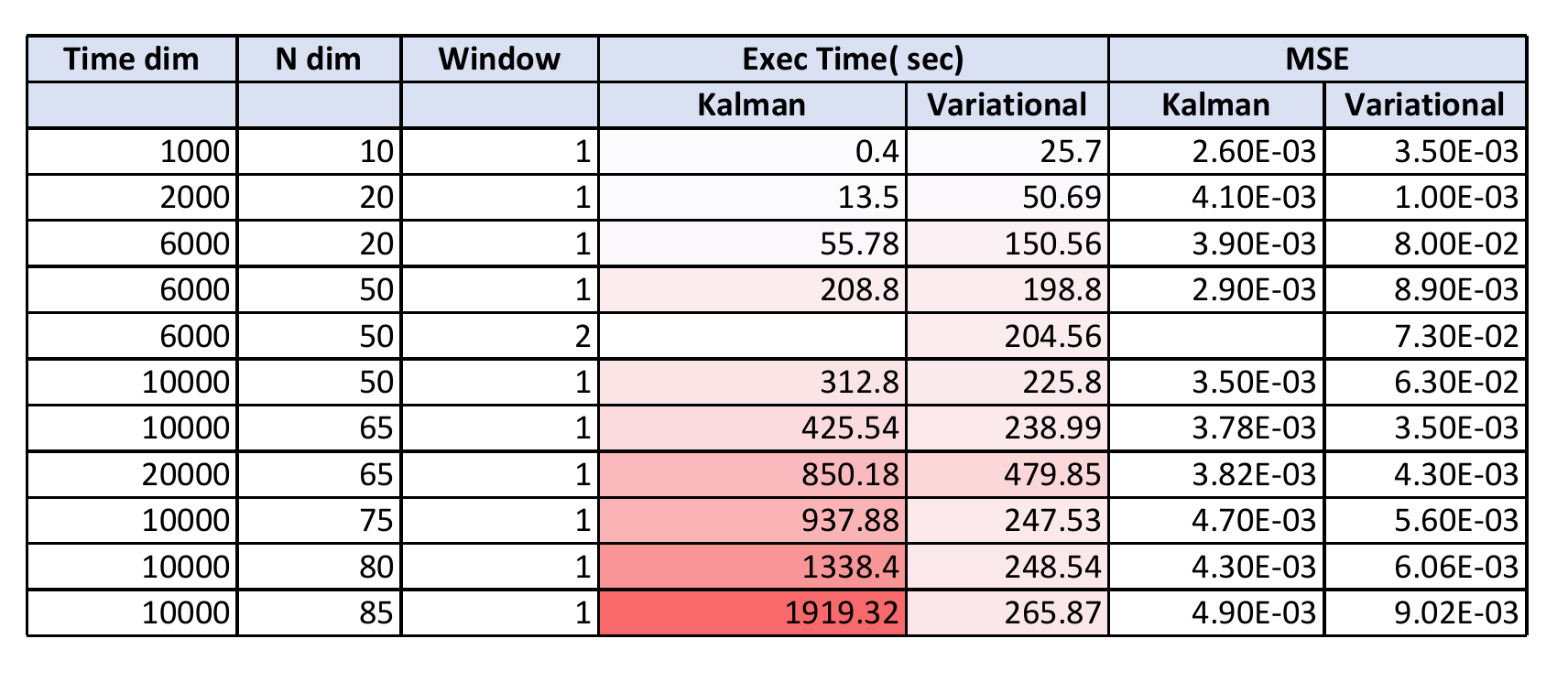}
\caption{Comparing Kalman vs Variational, with red gradient denoting the increasing execution time}
\label{benchmar}
\end{table}
We can observe that as the problem size (n dimension $\geq 50$) increases, the execution time increases for the Kalman by 10-40\%. In contrast, for variational, the increase is hardly 1\% for large problem size. With the increase in dimension, we see an increase in MSE for Variational, however the magnitude order ($10^{-3}$) remains the same, justifying the slight increase in MSE trade-off for lower execution time. \\ 
This can also be explained by looking at time complexities for the worst-case scenario. Time Complexity for matrix inversion (Kalman Gain) is about $O(n^3)$ (n is the size of observation - N dim) whereas for function minimisation using LBFGS is $O(mn)$ where m is the size of hessian history stored in memory, which is a constant hyper-param for LBFGS.\\
Another observation is that on comparing the experiments ran for both 4D-Var (window $>$ 1, represented by the entry that has a missing value for Kalman, window = 2)) and 3D-Var (window = 1), 4D-Var has longer execution time, suggesting difficulty to reach the minima. We believe the reason could be that adding future observations (as part of the summation of values in a window) in the cost function could lead to a non-convex function with many local minima. 
\vspace{-0.6em}
\subsection{Blockchain data}
\vspace{-0.6em}
After experimenting with well behaved TVP-VAR synthetic data, we empirically observed that as the dimension of data increases, the 3D-Var algorithm becomes more scalable; hence we use TVP-VAR-VI algorithm to interpolate latent parameters, $\beta$ for the actual blockchain datasets. \\
The dataset of interest is the ERC20 tokens being traded on Ethereum. We deal with mainly two data sources, ERC20 tokens on-chain data; and off-chain data from exchanges.

\textbf{On-Chain Data:} We identify on-chain data as trades on Ethereum, amount of ERC20 tokens traded between blockchain addresses, timestamp of the transaction, number of transactions etc.\\
\textbf{Off-Chain Data:} We refer to the off-chain data as open-closing prices and trading volumes obtained using https://min-api.cryptocompare.com per exchange and ERC20 token.

Applying this methodology to economic data can help us infer the state of the system, how some features affect/drive other features. The main features used for blockchain dataset analysis were amount sent to/received by exchanges, number of transactions sent or received, the closing price of the token being analysed and volume from or to the token being traded.\\
For example, we tried to visually interpret the effect of on-chain token inflow on returns and vice-versa for BNB token on Binance exchange. 
\begin{figure}[h!]
\centering
\begin{minipage}[t]{.48\textwidth}
\centering
\includegraphics[scale=0.2]{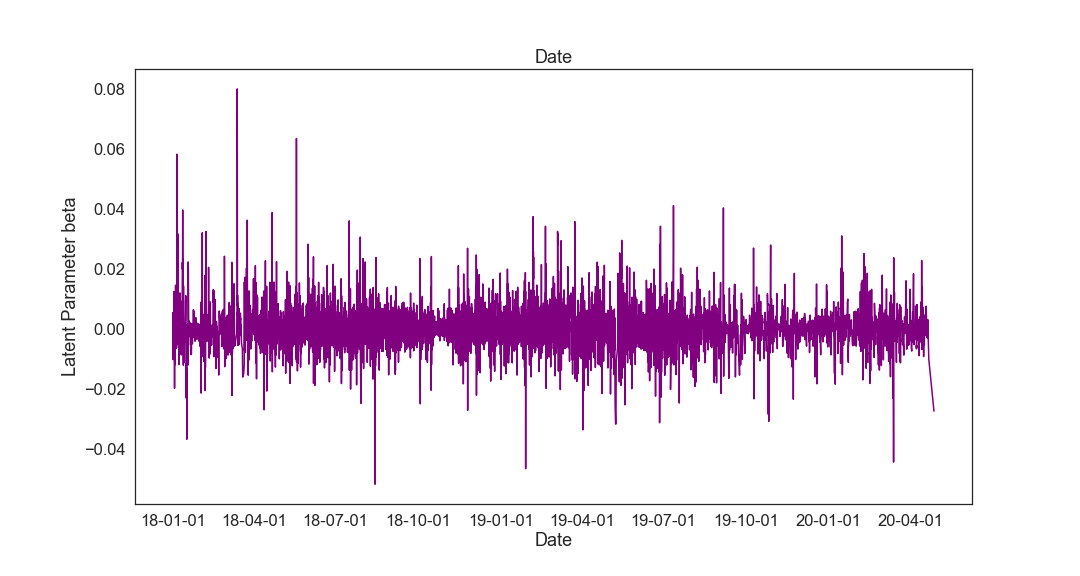}
\caption{Effect of amount to tokenflow on return for BNB token}
\label{fig:test1}
\end{minipage}\hfill
\begin{minipage}[t]{.48\textwidth}
\centering
\includegraphics[scale=0.2]{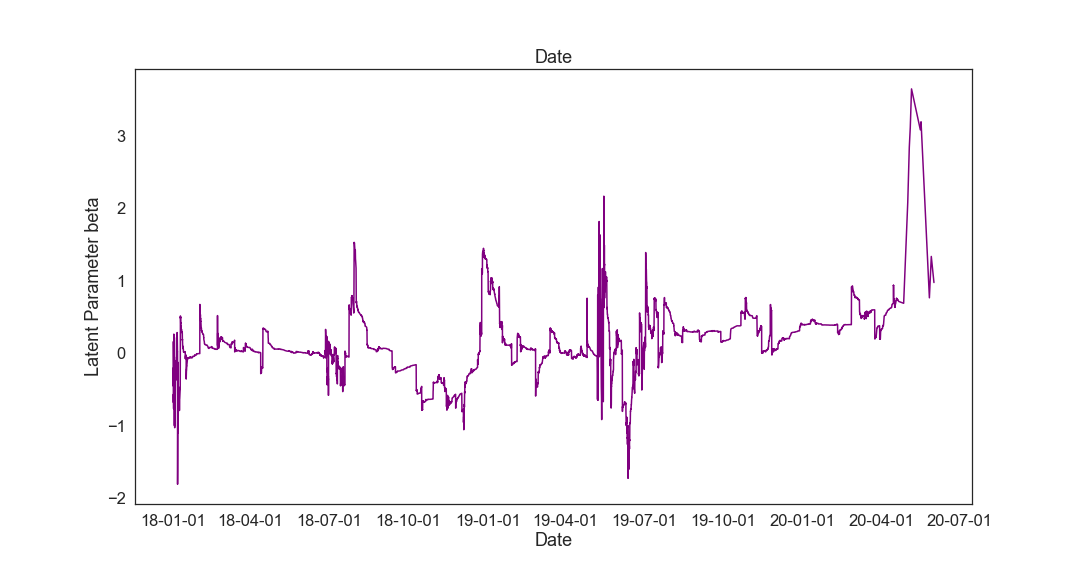}
\caption{Effect of return on token inflow for BNB Token}
\label{fig:BNBPricey}
\end{minipage}
\end{figure}
As follows from the Figure~\ref{fig:test1}, the dynamic is erratic and chaotic for on-chain effect on trading prices, leading to the suspicion that the interaction is not structural. Prominent theories exist for traditional exchanges describing that inflow drives the price of the asset. It is very likely not the case for Ethereum ecosystem as suggested by our interpretation of the dynamic.
However, we see a more prominent indication of the dynamic that represents the effect of returns on on-chain outflow/inflow of tokens when looking at Fig~\ref{fig:BNBPricey}. 
This implies a one-sided relationship, with token inflows having an erratic and weak relation on returns; whereas returns having a structural impact on the inflow. These results were also corroborated by experiments for other ERC20 tokens like OMG and TRX. \\
This is possible evidence that the price and trading action probably attracts attention and drives the inﬂow of tokens towards a particular exchange or it could be due to arbitrageurs who benefit from price difference and cause the fluctuation of tokenflow \cite{nadlerEconophysicalAnalysisBlockchain}. 
However, on-chain flows don't drive prices. It could be that over time the impact of on-chain activity has become more decoupled from price movements on exchanges. This can serve as an important indication factor to policymakers, which can help them to create and shape future regulations concerning ERC20 tokens. It shows, for example, price manipulation is unlikely to be driven by activities on the chain; thus, regulation can focus on exchange specific risk factors such as orderbook fraud or sentimental actions.
\section{Improving state-space modelling}
\vspace{-0.5em}
The state-space model stated in \eqref{state2_link} assumes that approximation of the beta evolution function, F is known. For the complex dynamics inherent in these economic systems, it is hard to know F, and with increasing dimensions, the non-linearities would become stronger; thus linear assumption of the forward model failing \cite{pawar2020long}.\\
The TVP-VAR-VI would be able to update the prior estimate of the latent parameter when observations are available; however, the state-space model would fail for out-of-sample predictions. \\
As seen in plots above, there have been fluctuations in values of $\beta$ that might not be driven only by the previous value and innovation but also a non-linear combination of some lagged values of $\beta$.\\
Running TVP-VAR-VI provides us with a complete state of time series of the latent parameter, $\beta$, with which we can obtain a surrogate forward model using machine learning tools.\\
Learning evolution of the unobserved $\beta$ parameter has been challenging. We were trying to predict with a time series which had high volatility and was chaotic in nature. Also, multivariate-time series forecasting has been challenging ongoing research as it needs to appropriately leverage the dynamics of multiple variables \cite{LSTNetlai2018modeling}. Hence, different architectures were tried before settling on the final model\vspace{-0.5em}.
\subsection{TVP-VARNet Model}
\begin{figure}[h!]
    \centering
    \includegraphics[scale=0.3]{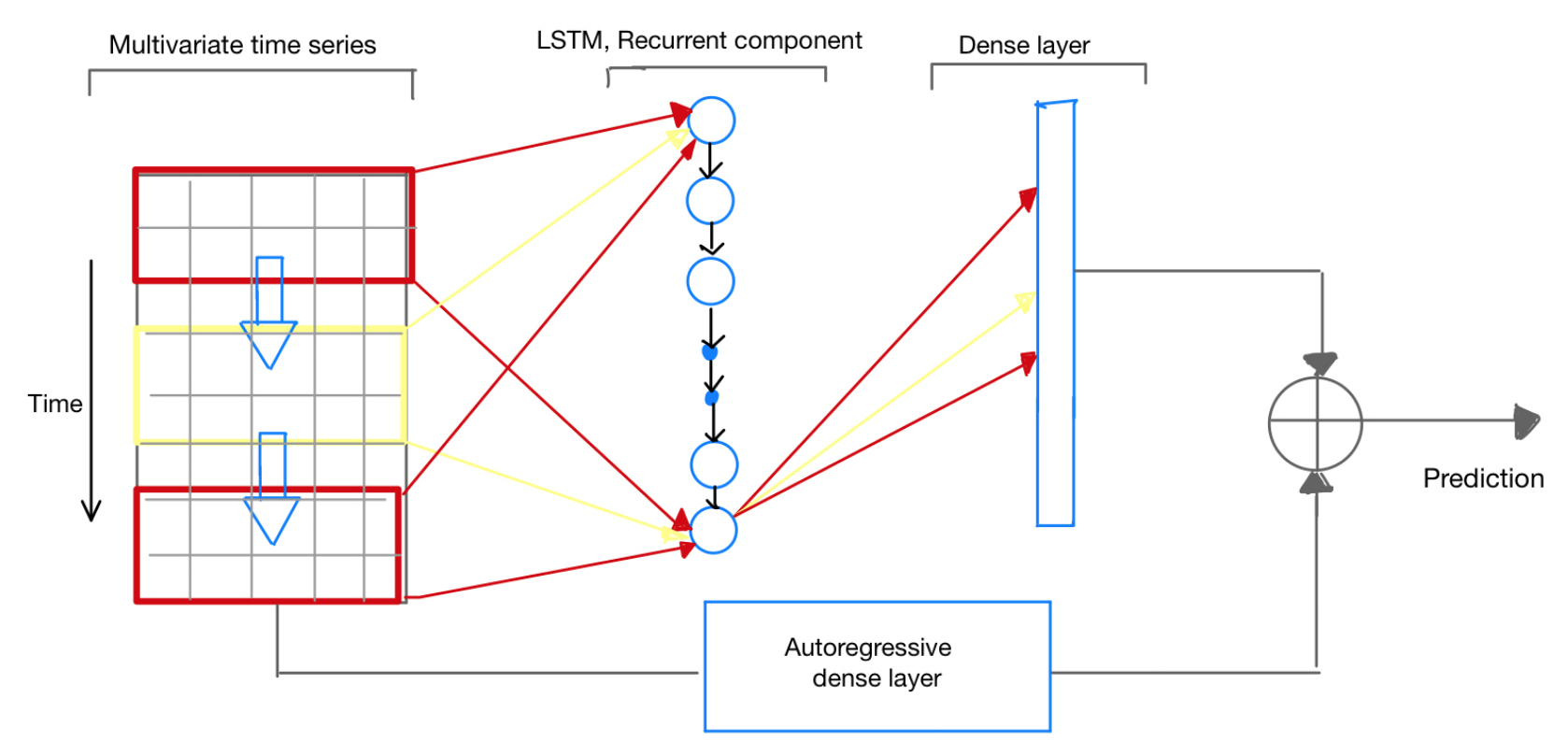}
    \caption{TVP-VARNet}
    \label{fig:nn.png}
\end{figure}

\begin{figure}[ht!]
\centering
\includegraphics[scale=0.3]{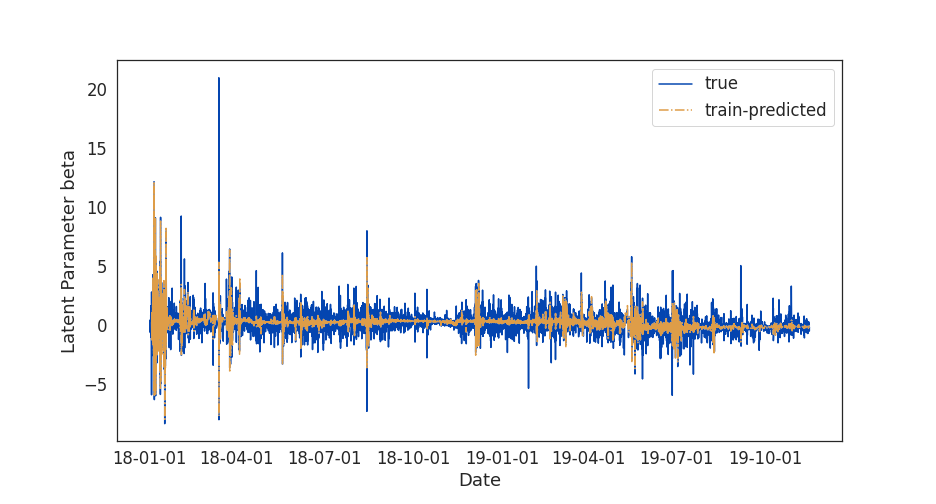}
\caption{LSTM model predictions for training and test set - fine grid for time steps}\label{fig:lstm} 
\end{figure}

\begin{figure}[ht!]
\centering
\includegraphics[scale=0.3]{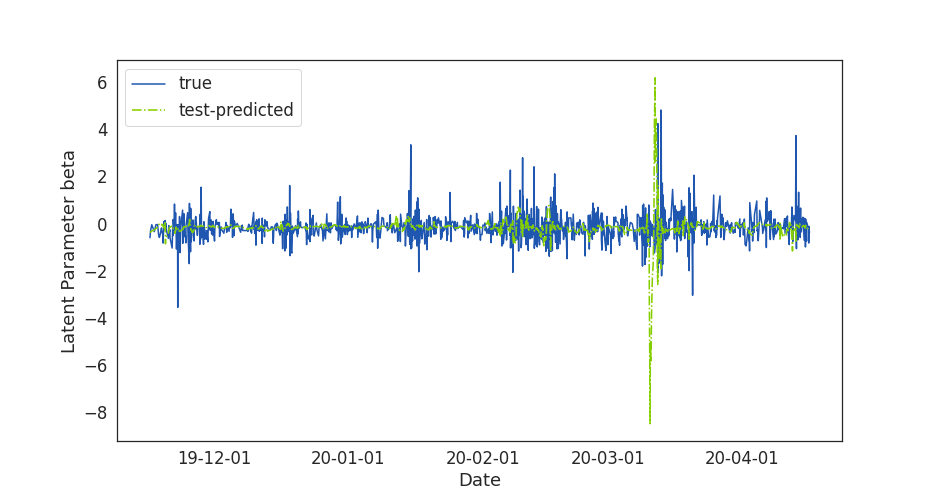}
\caption{LSTM model predictions for training and test set - course grid for time steps}\label{fig:lstm2} 
\end{figure}

\begin{figure}[h!]
\centering
\includegraphics[scale=0.3]{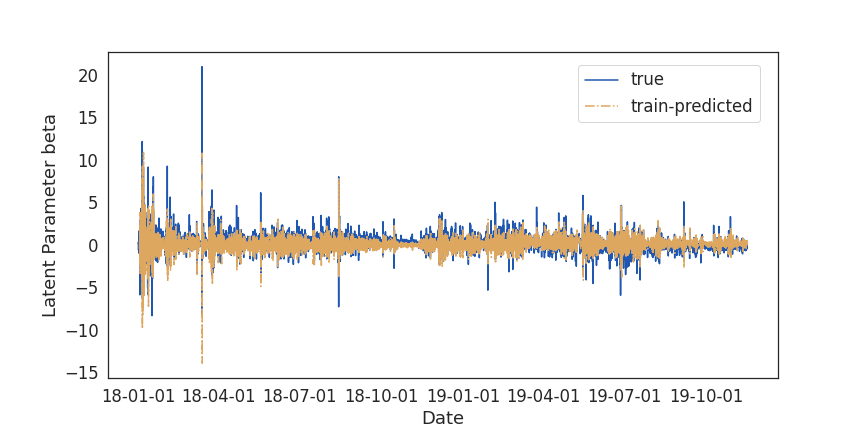}
\caption{TVP-VARNet model one-step predictions for training set} 
\label{fig:ml} 
\end{figure}

\begin{figure}[h!]
\centering
\includegraphics[scale=0.3]{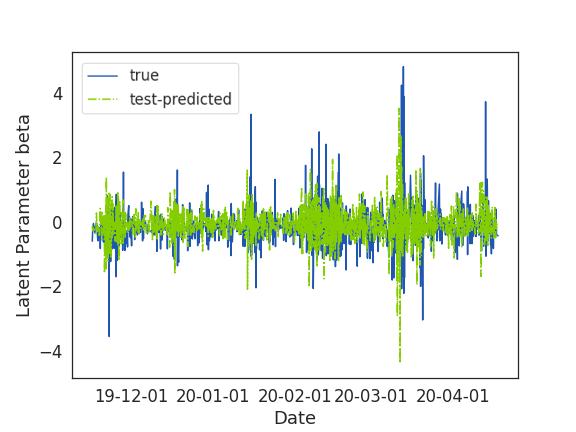}
\caption{TVP-VARNet model one-step predictions for  test set} 
\label{fig:ml2} 
\end{figure}

\vspace{-0.6em}
Figure~\ref{fig:nn.png} presents an overview of our proposed architecture, TVP-VARNet to model latent parameter dynamics of blockchain data.
The architecture benefits from an ensemble of two neural networks, an LSTM and an Autoregressive dense network with their outputs combined. The LSTMs has been successful due to its ability to capture and effectively exploit temporal dependencies, and take history into account for future state prediction.\\
However, as suspected by \citet{LSTNetlai2018modeling}, LSTM could lead to an issue where the scale of outputs might not be sensitive to the scale of inputs. In the evolution of our unobserved $\beta$ parameter, we see constant changes which are also in a non-periodic manner, resulting in poor forecasting of the model with only LSTM. To address this issue, similar in spirit to the LSTNet \cite{LSTNetlai2018modeling}, we
combine the final prediction of LSTM with
a linear part, hence the model benefits from non-linear part having recurrent patterns through LSTM and a linear part through this component. By plotting the predictions on training and test set using only LSTM (see Figures~\ref{fig:lstm}-\ref{fig:lstm2}) and using the TVP-VARNet (see Figures~\ref{fig:ml}-\ref{fig:ml2}), we can understand the cruciality of the Autoregressive component.

TVP-VARNet does a good job in generalising, by working appropriately for both kinds of evolution, chaotic and structured. Our architecture was able to adapt and forecast a satisfactory result for two-step ahead predictions.\\
It is important to note that the combination of TVP-VAR-VI and TVP-VARNet is crucial. TVP-VAR-VI provides us with a set of latent parameters to train our model, whereas TVP-VARNet helps us with the out-of-sample forecast. As and when new observations are available, TVP-VAR-VI can be run to obtain updated $\beta$ values to ﬁne tune the TVP-VARNet hence giving us the ability to bring continuous correction to our machine learning model.

\vspace{-0.7em}
\section{Conclusion, future work and social impact}
\vspace{-0.8em}
The findings of this study could be understood as further validation of treating the cryptocurrency system as an econophysical system. By ingesting data from various on-chain transactions and trading information, we perform Bayesian inference to infer the latent time-series, which can significantly help in analysing the dynamics that exist in the economic system. We built a scalable novel TVP-VAR-VI algorithm. Before TVP-VAR-VI, the Kalman Filter approach was computationally expensive. We were able to reduce the execution time in the range of 10-40\% as dimensions increased. \\
Our interpretation and results of the unobserved state are broadly consistent with conclusions drawn by Nadler et.al. The most prominent finding to emerge was the direction of interaction between on-chain token transaction/flows, and off-chain trading action was one-sided. Token inflow’s effect on trading actions resulted in weak and chaotic interaction; however, has a high-persistent effect when reversing the roles. Trading price action has a considerate impact on driving the inflow of tokens for an exchange. We also observed that change in trading volume also had a structural effect on the returns.\\
One of the other significant contributions is challenging the latent state-space evolution and replacing it with a more nuanced, non-linear neural network model. The TVP-VARNet architecture models both kinds of latent dynamics: chaotic and structural. The relevance of having a surrogate forward model for the latent parameter evolution is supported by the fact that the out-of-sample forecast faces many hurdles with the simplistic state-space model. Our TVP-VARNet brings the possibility to forecast latent parameters for certain future time-steps effectively.\\
One of the directions for future work includes further analysis of this dynamic on highly granular order-book data. Ideally, these experiments can be replicated to any system that can be modelled as a TVP-VAR model, to understand their latent dynamics in a scalable fashion.

\textbf{Social Impact} The insight to the time-varying dynamical interrelationship between tokenflow, price movements and volume activities can give much-needed information about price and market dynamics in the blockchain ecosystem, which has been unexplored and unregulated. We believe that the traditional financial market rules do not apply to a peer-to-peer based financial market, hence can pave the way for blockchain analyst and policymakers to understand what factors might or might not affect said movements. This intern can help in avoiding fraudulent activities or arbitrage. Our work can we extended to any econophysical system modelled using TVP-VAR equations.

\bibliography{references}
\end{document}